\documentstyle[aps,prl]{revtex}

\begin{document}

\title{A Two Parameter Four Neutrino Mixing Model with an Exchange
  Symmetry of the Mass Doublet Neutrinos}

\author{E.M. Lipmanov}

\address{40 Wallingford Rd. \#272, Brighton, MA 02135, USA}

\maketitle
 
\begin{abstract}
In a model with four neutrino mass eigenstates grouping in two narrow
doublets the condition of maximal mixing of the doublet neutrino
components leads to a simple 4-neutrino mixing matrix which is related
to two connected effective two-neutrino mixing patterns.  As a result,
large values of the neutrino oscillation amplitudes for both the solar
and atmospheric neutrino anomalies appear quite naturally if the LSND
data are accepted.
\end{abstract}





In the course of the last few years it was gradually shown
\cite{numodels,bilenky} that all the currently available data on neutrino
oscillations can be described in a scheme with four neutrino mass states 
where the four masses are divided into two pairs of close masses
separated by a gap of the order of 1~eV.  The data of the solar,
atmospheric, and LSND experiments can be explained if there is neutrino
mixing with the following values of the neutrino mass differences
\cite{bilenky}:
\begin{eqnarray}
  \label{eq:bilenky}
  \Delta m_1^2 & \equiv & \Delta m^2_{solar} \sim 10^{-10}(Vac),~{\rm
    or} \sim 10^{-5}(MSW)~{\rm eV}^2,\nonumber\\
  \Delta m_2^2 & \equiv & \Delta m^2_{atm} \sim 10^{-3} {\Large -}
  10^{-2}~{\rm eV}^2,\\
  \Delta m_{12}^2 & \equiv & \Delta m^2_{LSND} \sim 1~{\rm eV}^2.\nonumber
\end{eqnarray}
Only two possible schemes of the neutrino mass spectra are acceptable,
(\ref{eq:doublet}) and (B), with 
\renewcommand{\theequation}{\Alph{equation}}\setcounter{equation}{0}
\begin{equation}
  \label{eq:doublet}
  \underbrace{\overbrace{m_1 < m_1^\prime}^{atm} \ll 
    \overbrace{m_2 < m_2^\prime}^{solar}}_{LSND}~,
\end{equation}
\renewcommand{\theequation}{\arabic{equation}}\setcounter{equation}{1}
and in scheme (B) the positions of the ``solar'' and ``atm'' doublet
splittings are exchanged.

In the present paper a methodological attempt is made to unify the
solar, atmospheric and the LSND data, and also the reactor and
accelerator oscillation data, in a simple explicit two parameter
4-neutrino mixing model with two mass doublets and maximum neutrino
doublet mixing with only one parameter to adjust (the $\theta_{LSND}$
angle) for the currently available data.  Let us denote by $\nu_1,
\nu_1^\prime, \nu_2$ and $\nu_2^\prime$, the four neutrino mass
eigenstates.  Then the four eigenstates of the exchange symmetry are
\begin{equation}
  \label{eq:4eigen}
  \nu^{s,a}_{1,2} = \frac{1}{\sqrt{2}}(\nu_{1,2} \pm \nu_{1,2}^\prime).
\end{equation}
The physical meaning of the Eq.~(\ref{eq:4eigen}) is that the neutrino
mass states within the two doublets are mixed maximally, which seems a
natural suggestion for the very close neutrino mass states in the weak
interactions.  The four phenomenological neutrino states -- the three
weak interaction eigenstates plus one sterile neutrino $\nu_s$ -- can be
expressed in the form:
\begin{eqnarray}
  \label{eq:nue}
  \nu_e & = & ~~\nu_1^s \cos \theta + \nu_2^s \sin \theta,\\
  \label{eq:numu}
  \nu_\mu & = & -\nu_1^s \sin \theta + \nu_2^s \cos \theta,\\
  \label{eq:nutau}
  \nu_\tau & = & ~~\nu_1^a \cos \bar{\theta} + \nu_2^a \sin \bar{\theta},\\
  \label{eq:nus}
  \nu_s & = & -\nu_1^a \sin \bar{\theta} + \nu_2^a \cos \bar{\theta}.
\end{eqnarray}
Any permutations of the superscripts $s$ and $a$ in the
Eqs.~(\ref{eq:nue})--(\ref{eq:nus}) will not change the results if only
the $\nu_e$ and $\nu_\mu$ in the Eqs.~(\ref{eq:nue}) and
(\ref{eq:numu}) share the same two exchange eigenstates.  Note that all
found in Ref.~\cite{bilenky} criteria for a 4-neutrino two mass doublets
mixing scheme to be viable are fulfilled in the explicit model
(\ref{eq:nue})--(\ref{eq:nus}) with $\theta = \theta_{LSND}$.

The probability of the $\nu_\mu \rightarrow \nu_e$ oscillations which
are relevant to the LSND effect is
\begin{equation}
  \label{eq:lsnd_prob}
  \left | \left \langle \nu_\mu(0) | \nu_e(L) \right \rangle \right |^2
  = \sin^22\theta \left [ \left \langle \! \! \! \left \langle \sin^2 
        \left ( \frac{\Delta m^2_{12}L}{4E}
        \right ) \right \rangle \! \! \!
    \right \rangle - \frac{1}{4}\sin^2\left ( \frac{\Delta m_1^2 L}{4E} 
    \right ) - \frac{1}{4}\sin^2\left ( \frac{\Delta m_2^2 L}{4E} 
    \right ) \right ],
\end{equation}
where $\theta = \theta_{LSND}$, and the symbol $\langle \! \langle
~ \rangle \! \rangle$ in the first term denotes the arithmetic
mean value of the appropriate four factors related to the four large
mass squared differences among the two neutrino mass doublets.

The notations are
\begin{equation}
  \label{eq:notation}
  \Delta m^2_{1,2} = m^2_{1,2} - m^{\prime 2}_{1,2}~,\;\; \Delta m^2_{12} 
  \cong m^2_2 - m^2_1~,
\end{equation}
$E$ is the initial beam energy and $L$ is the distance from the source.
The first term in Eq.~(\ref{eq:lsnd_prob}) describes the LSND effect
\cite{lsnd}:
\begin{equation}
  \label{eq:lsnd_effect}
  \sin^22\theta \approx 10^{-2}.
\end{equation}
The second and third terms in Eq.~(\ref{eq:lsnd_prob}) give respectively
small $\nu_e\rightarrow\nu_\mu$ and $\nu_\mu\rightarrow\nu_e$
contributions to the dominant terms in the Eqs.~(\ref{eq:nue_app}) and
(\ref{eq:w}) for the solar and atmospheric oscillation probabilities.

The probability of the total appearance oscillations of the electron
neutrino into all the other neutrino states $(\nu_\mu+\nu_\tau+\nu_s)$:
\begin{equation}
  \label{eq:nue_app}
  W(\nu_e\rightarrow\nu_\mu,\nu_\tau,\nu_s) \cong \cos^4\theta\sin^2 
  \left ( \frac{\Delta m_1^2 L}{4 E} \right ) + \sin^4\theta\sin^2 \left (
    \frac {\Delta m_2^2 L}{4 E} \right ) + \sin^22\theta\sin^2 \left (
    \frac{\Delta m^2_{12} L}{4 E} \right ),
\end{equation}
where we disregarded the doublet widths in the third term in comparison
with the doublet separation.  The first term in Eq.~(\ref{eq:nue_app})
is the dominant one; it describes the solar neutrino oscillations with
the amplitude value
\begin{equation}
  \label{eq:a_solar}
  A_{solar} \cong \cos^4 \theta \cong 1
\end{equation}
It does not depend on the angle $\bar{\theta}$ in the model
(\ref{eq:nue})--(\ref{eq:nus}) and is fully determined by the LSND data
\cite{lsnd}.  Because the $\sin^2\theta_{LSND}$ is very small, the solar 
vacuum neutrino disappearance oscillations in the model
(\ref{eq:nue})--(\ref{eq:nus})  are mostly due to the transitions
$\nu_e\rightarrow\nu_\tau+\nu_s$, 
\begin{equation}
  \label{eq:w_sol}
  W(\nu_e\rightarrow\nu_e) \cong 1 -
  \cos^2\theta(\cos^2\bar{\theta}+\sin^2\bar{\theta})\sin^2 \left (
    \frac{\Delta m_1^2 L_I}{4 E} \right ) \cong \cos^2 \left (
    \frac{\Delta m_1^2 L_I}{4 E} \right ).
\end{equation}


This is just the original still viable Pontecorvo~\cite{pontecorvo}
2-neutrino maximal mixing solution for the solar neutrino deficit (here
$L_I$ is the Earth-Sun distance). In Eqs.~(\ref{eq:nue_app}),
(\ref{eq:w_sol}), and (\ref{eq:w}), a special feature of the model
(\ref{eq:nue})--(\ref{eq:nus}) is reflected: the appearance oscillations
between the electron and muon neutrinos $(\nu_e, \nu_\mu)$, on the one
hand, and the tau and sterile neutrinos $(\nu_\tau, \nu_{s})$ on the
other hand, are determined by the ``small'' doublet mass squared
differences $\Delta m_1^2$ and $\Delta m_2^2$ with only small
corrections from the ``large'' neutrino mass squared difference $\Delta
m_{12}^2$ seen in the LSND experiment. This means that the model
predicts negative results with $\sin^2 2\theta_{\mu\tau} \ll \sin^2
2\theta_{LSND}$ for the short baseline $\nu_\mu$ oscillation experiments
such as CHORUS~\cite{chorus}.

The probability of the total appearance oscillations of the muon
neutrinos into all the other neutrino states $(\nu_e, \nu_\tau,
\nu_s)$ is:
\begin{equation}
  W(\nu_\mu \rightarrow \nu_e, \nu_\tau, \nu_s) \cong 
  \cos^4 \theta \sin^2 \left ( \frac{\Delta m_2^2 L}{4 E} \right ) +
  \sin^4 \theta \sin^2 \left ( \frac{\Delta m_1^2 L}{4 E} \right ) + 
  \sin^2 2\theta \sin^2 \left ( \frac {\Delta m_{12}^2 L }{4 E} \right ).
  \label{eq:w}
\end{equation}
The first term in Eq.~(\ref{eq:w}) is the dominant one, and describes
the atmospheric neutrino oscillations with the amplitude value
\begin{equation}
  A_{atm} \cong A_{solar} \cong 1.
  \label{eq:amplitudes}
\end{equation}
From the Eqs.~(\ref{eq:lsnd_prob}),(\ref{eq:nue_app}),(\ref{eq:w}) and
the LSND data it follows that in the model (\ref{eq:nue})--(\ref{eq:nus})
the solar and atmospheric neutrino deficits are dominated by the
transitions of the electron and muon neutrinos into the tau and sterile
neutrinos in a complementary way: if the solar $\nu_e$ disappearance is
mostly $\nu_e \rightarrow \nu_\tau$ dependent (with $\bar{\theta} <
\pi/4$), then the atmospheric $\nu_\mu$ disappearance is mostly $\nu_\mu
\rightarrow \nu_s$ dependent, and vice versa (with $\bar{\theta} >
\pi/4$).  The second possibility is preferred by the latest
Super-Kamiokande data \cite{superk} which favor the appearance
$\nu_\mu\rightarrow \nu_\tau$ oscillations with large amplitude $\sin^2
2\theta \simeq 1$ as a better match to the atmospheric $\nu_\mu$ deficit
data.

From the Eq.~(\ref{eq:nue_app}) it follows that the deviation from unity
of the $\nu_e \rightarrow \nu_e$ survival probability is small, $\lesssim
\sin^2 2\theta_{LSND}$, for oscillation distances $L \ll 4\pi E/\Delta
m_1^2$. This is in good agreement with the $\nu_e$ disappearance
oscillation data such as Bugey, and the appearance oscillation $\nu_\mu
\rightarrow \nu_e$ data from BNL E776 experiment~\cite{bugeybnl}, and
also with the long-baseline ($L=1$ km) reactor $\nu_e$-disappearance
CHOOZ experiment \cite{chooz} where this restriction on the distances is
fulfilled quite well.

From the Eq.~(\ref{eq:w}) it follows that the deviation from unity of
the $\nu_\mu\rightarrow\nu_\mu$ survival probability will be small only
at distances $L \ll 4\pi E/\Delta m^2_2$, and so a discovery of a large
$E/L$ dependent $\nu_\mu$-disappearance could be within the reach of the
coming long baseline $\nu_\mu$ oscillation experiments.

With regard to the solar $\nu_e$-disappearance data, the large
oscillation amplitude (\ref{eq:a_solar}) is in disagreement with the
small angle MSW solution, but not so with the other two solutions: the
large angle MSW solution and the vacuum oscillation solution which are
compatible with large appearance oscillation amplitudes (see the
discussion in Ref.~\cite{krastev}).

In conclusion, a methodologically simple explicit minimal parameters
4-neutrino mixing model in a two mass doublet scheme with maximal
doublet neutrino mixing is considered. The main results are: 1) the
equality Eq.~(\ref{eq:amplitudes}) of the naturally large solar and
atmospheric dominant disappearance oscillation amplitudes; 2) the
prevailing contributions of the $\nu_e \rightarrow \nu_\tau + \nu_s$ and
$\nu_\mu \rightarrow \nu_\tau + \nu_s$ transitions in the solar and
atmospheric appearance oscillations in a complementary way; 3) the model
unifies naturally the three currently available positive neutrino
oscillation data from the solar and atmospheric neutrino deficits and
the LSND experiment, plus the negative reactor and accelerator
oscillation data, with only one parameter to adjust -- the $\theta =
\theta_{LSND}$ angle; 4) the neutrino doublet exchange symmetry seems
methodologically useful; maximal mixing in the weak interactions is an
attractive suggestion for very close neutrino states in the mass
doublets; 5) The present model appears to be in a remarkable agreement
with the positive data on atmospheric $\nu_\mu$ oscillations reported by
the Super--Kamiokande group~\cite{sknu98}. As I learned after this work
was finished, the results here which are relevant to the
Super--Kamiokande data are in good agreement with the those obtained in
a recent comprehensive study of the neutrino data in Ref.\cite{barger}.
The approach here is different, however, it is deductive instead: one
physical hypothesis -- maximal neutrino doublet mixing in a
four-neutrino model -- leads necessarily to all the predictions if the
LSND data are accepted.

I thank E.~Kearns for a preprint of his paper.

\end{document}